\documentclass
[final,10pt,letterpaper,twocolumn,lengthcheck,showpacs,prl]{revtex4}%
\usepackage{graphicx}%
\usepackage{amsmath}%
\usepackage{amsfonts}%
\usepackage{amssymb}
\begin{document}
\title{Deterministic and Highly Efficient Quantum Cryptography with Entangled Photon Pairs}
\author{Zhi Zhao, Tao Yang, Zeng-Bing Chen, Jiangfeng Du, and Jian-Wei Pan}
\affiliation{Department of Modern Physics, University of Science and Technology of China,
Hefei, Anhui 230027, People's Republic of China}
\pacs{03.67.Lx, 42.50.Ar}

\begin{abstract}
We present a protocol for deterministic and highly efficient quantum
cryptography with entangled photon pairs in a 4$\times4$-dimentional Hilbert
space. Two communicating parties, Alice and Bob first share a both
polarization- and path-entangled photon pair, and then each performs a
complete Bell-state measurement on their own photon in one of two
complementary Bell-state bases. It is demonstrated that each measurement in
which both Alice and Bob register a photon can build certain perfect
correlation and generate $1.5$ key bits on average. The security of our
protocol is guaranteed by the non-cloning theorem.

\end{abstract}
\maketitle

The central task for quantum cryptography is to establish secure keys before
the transmission of message between two parties \cite{gis}. In recent years,
both BB84 \cite{ben1} and Ekert91 \cite{ekert} cryptographic protocols have
been successfully demonstrated in numerous experiments
\cite{butter1,jen,naik,tit,butter2,kurt}. However, one drawback of the above
schemes is the non-deterministic feature, that is, only less than 50\% of all
detected qubits can be further used as key bits. This is a serious practical
problem, because in the one-time-pad secret-key cryptosystem, the length of
secret key must be the same as the ciphertext \cite{sch}. Although some
deterministic cryptographic schemes based on orthogonal states have been
proposed recently \cite{Gold,koashi,cabello1}, the long storage rings
required, which is an essential ingredient of those schemes, lead to a low
efficiency for the transmission of polarized photons. More recently,
deterministic quantum cryptography and secret direct communication have also
been proposed with single-photon two-qubit states \cite{beige1}. By exploiting
the states in four- or more dimensional Hilbert space, each transmitted photon
can be used as a key bit with the help of classical communication;
eavesdroppers can be detected within the scheme. However, every photon
detected can establish only one key bit.

In this work, we propose a deterministic and highly efficient quantum
cryptography protocol using doubly entangled photon pairs
\cite{Simon-Pan,chen}. In the scheme, two communicating parties, Alice and
Bob, first share a polarization- and path-entangled photon pair, then each
performs a complete Bell measurement on Alice's (Bob's) photon in one of two
complementary Bell bases. In this way, they can establish $1.5$ key bits on
average for each measurement where both Alice and Bob register a single
photon. We also show that the security of our protocol is guaranteed by the
non-cloning theorem.

In order to establish a key bit at each run - that is - in a determinstic way,
it would require that, on the one hand, each assigned joint measurement must
generate certain perfect correlations between two communicating parties; and
on the other hand, some of these perfect correlations are sufficient to
perform the validation of eavesdropper. Our quantum cryptography protocol is
based on this observation. Although it is not possible to contruct such a
protocol for two observers with only one Einstein-Poldolsky-Rosen (EPR) pair,
one can indeed achieve this goal using two EPR pairs, provided that one can
achieve the controlled-NOT operation between two indepedent photons
\cite{cabello2}.

In our scheme, instead of utilizing two EPR pairs, we make use of only one
photon pair where the two photons are entangled both in polarization and in
path degree of freedom \cite{chen}. In contrast to the two pair scheme, there
are two obvious advantages in our single pair scheme. First, the required
complete Bell-state measurement which projects Alice's (Bob's) photon onto a
polarization-path entangled state in one of two complementary bases is
feasible with only linear optics and, actually, has been experimentally
implemented \cite{martini}. Second, using two usual polarization-entangled
photon pairs will lead to very low efficiency due to the very low probability
of simultaneously generating two entangled photon pairs. Our scheme is
deterministic in the sense that all measurements in which both Alice and Bob
register a photon can be used to establish perfect correlations, which can
then be converted into key bits and a few of them are enough to perform the
detection of any eavesdroppers. Entanglement exploited in the present scheme
is used both to provide security against eavesdropping and to enhance the key
bit generation.

We now explain how our scheme works. As shown in Fig. $1$, two communicating
parties, Alice and Bob, first share a both polarization- and path-entangled
photon pair. It can be generated using parametric down conversion (PDC) by the
method described in Ref. \cite{pan2}. As explained in refs.
\cite{Simon-Pan,chen}, after the UV laser passes through the BBO crystal
twice, the state corresponding to the case where there is one and only one
pair production is given by%

\begin{align}
\left\vert \Psi\right\rangle _{A1B2}  & =\frac{1}{2}\left(  \left\vert
H\right\rangle _{1}\left\vert V\right\rangle _{2}-\left\vert V\right\rangle
_{1}\left\vert H\right\rangle _{2}\right)  \nonumber\\
& \otimes\left(  \left\vert a_{1}\right\rangle \left\vert b_{2}\right\rangle
-\left\vert b_{1}\right\rangle \left\vert a_{2}\right\rangle \right)
.\label{state}%
\end{align}
\ \ \ \ \ \ \ \ \ \ \ \ \ \ \ \ \ \ \ \ \ \ \ \ \ \ \ \ \ \ \ \ \ \ \ \ \ \ \ \ \ \ \ \ \ \ \ \ \ \ \ \ \ \ \ \ \ \ \ \ \ \ \ \ \ \ \ \ \ \ \ \ \ \ \ \ \ \ \ \ \ \ \ \ \ \ \ \ \ \ \ \ \ \ \ \ \ \ \ \ \ \ \ \ \ \ \ \ \ \ \ \ \ \ \ \ \ \ \ \ \ \ \ \ \ \ \ \ \ \ \ \ \ \ \ \ \ \ \ \ \ \ \ \ \ \ \ \ \ \ \ \ \ \ \ \ \ \ \ \ \ \ \ \ \ \ \ \ \ \ \ \ \ \ \ \ \ \ \ \ \ \ \ \ \ \ \ \ \ \ \ \ \ \ \ \ \ \ \ \ \ \ \ \ \ \ \ \ \ \ \ \ \ \ \ \ \ \ \ \ \ \ \ \ \ \ \ \ \ \ \ \ \ \ \ \ \ \ \ \ \ \ \ \ \ \ \ \ \ \ \ \ \ \ \ \ \ \ \ \ \ \ \ \ \ \ \ \ \ \ \ \ \ \ \ \ \ \ \ \ \ \ \ \ \ \ \ \ \ \ \ \ \ \ \ \ \ \ \ \ \ \ \ \ \ \ \ \ \ \ \ \ \ \ \ \ \ \ \ \ \ \ \ \ \ \ \ \ \ \ \ \ \ \ \ \ \ \ \ \ \ \ \ \ \ \ \ \ \ \ \ \ \ \ \ \ \ \ \ \ \ \ \ \ \ \ \ \ \ \ \ \ \ \ \ \ \ \ \ \ \ \ \ \ \ \ \ \ \ \ \ \ \ \ \ \ \ \ \ \ \ \ \ \ \ \ \ \ \ \ \ \ \ \ \ \ \ \ \ \ \ \ \ \ \ \ \ \ \ \ \ \ \ \ \ \ \ \ \ \ \ \ \ \ \ \ \ \ \ \ \ \ \ \ \ \ \ \ \ \ \ \ \ \ \ \ \ \ \ \ \ \ \ \ \ \ \ \ \ \ \ \ \ \ \ \ \ \ \ \ \ \ \ \ \ \ \ \ \ \ \ \ \ \ \ \ \ \ \ \ \ \ \ \ \ \ \ \ \ \ \ \ \ \ \ \ \ \ \ \ \ \ \ \ \ \ \ \ \ \ \ \ \ \ \ \ \ \ \ \ \ \ \ \ \ \ \ \ \ \ \ \ \ \ \ \ \ \ \ \ \ \ \ \ \ \ \ \ \ \ \ \ \ \ \ \ \ \ \ \ \ \ \ \ \ \ \ \ \ \ \ \ \ \ Here
$\left\vert H\right\rangle _{1\text{ }}$and $\left\vert V\right\rangle _{1}$
represent the polarization states of photon $1$ in path $a_{1}$ or $b_{1}$,
and $\left\vert H\right\rangle _{2\text{ }}$and $\left\vert V\right\rangle
_{2}$ represent the states of photon $2$ in path $a_{2}$ or $b_{2}$. Then let
photon $1$ in path $a_{1}$ and $b_{1}$ move towards Alice and photon $2$ in
path $a_{2}$ and $b_{2}$ towards Bob. With such state Alice and Bob can
perform entanglement swapping \cite{zu,pan1} or all-versus-nothing\ test of
quantum mechanics against local realism \cite{chen,cabello2}. In the
following, we are going to show how entanglement swapping of the quantum state
$\left\vert \Psi\right\rangle _{A1B2}$ provides us a deterministic and
efficient quantum cryptography.

\begin{figure}[tb]
\includegraphics[width=\columnwidth]{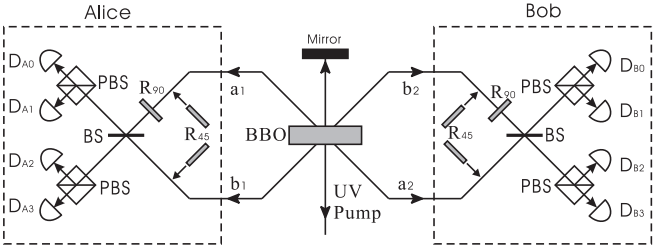}\caption{Schematic diagram of
generating both polarization- and path-entangled photon pairs and performing
the complete Bell measurement in one of the two complentary bases. R$_{90}$
and R$_{45}$ are half-wave plates that rotating the polarization of photons by
90$^{0}$ and 45$^{0}$ respectively. When inserting R$_{90}$ into the mode
a$_{1}$ (or b$_{2}$), Alice (or Bob) can perform the complete Bell-state
measurement on the state of $\left\{  \left\vert \Phi^{\pm}\right\rangle
_{ij},\left\vert \Psi^{\pm}\right\rangle _{ij}\right\}  $. When inserting two
45$^{0}$s into the modes a$_{1}$ and b$_{1}$ (or a$_{2}$ and b$_{2}$)
simultaneously, Alice (or Bob) can perform the complete Bell-state measurement
on the state of $\left\{  \left\vert \chi^{\pm}\right\rangle _{ij},\left\vert
\omega^{\pm}\right\rangle _{ij}\right\}  $\cite{martini}.}%
\end{figure}

In order to establish secure key bits between Alice and Bob, they can
\textit{randomly} choose one of two complementary Bell-state bases $\left\{
\left\vert \Phi^{\pm}\right\rangle _{ij},\left\vert \Psi^{\pm}\right\rangle
_{ij}\right\}  $ (type-I) or $\left\{  \left\vert \chi^{\pm}\right\rangle
_{ij},\left\vert \omega^{\pm}\right\rangle _{ij}\right\}  $ (type-II) $(ij=A1$
or $B2)$ and then perform Bell-state measurement. The two Bell-state bases are
defined as following:
\begin{equation}
\left\{
\begin{array}
[c]{c}%
\left\vert \Phi^{\pm}\right\rangle _{ij}=\frac{1}{\sqrt{2}}\left(  \left\vert
H\right\rangle _{i}\left\vert a_{j}\right\rangle \pm\left\vert V\right\rangle
_{i}\left\vert b_{j}\right\rangle \right) \\
\left\vert \Psi^{\pm}\right\rangle _{ij}=\frac{1}{\sqrt{2}}\left(  \left\vert
H\right\rangle _{i}\left\vert b_{j}\right\rangle \pm\left\vert V\right\rangle
_{i}\left\vert a_{j}\right\rangle \right)
\end{array}
,\right. \label{type1}%
\end{equation}
and
\begin{equation}
\left\{
\begin{array}
[c]{c}%
\left\vert \chi^{\pm}\right\rangle _{ij}=\frac{1}{2}\left[  \left\vert
H\right\rangle _{i}\left(  \left\vert a_{j}\right\rangle +\left\vert
b_{j}\right\rangle \right)  \pm\left\vert V\right\rangle _{i}\left(
\left\vert a_{j}\right\rangle -\left\vert b_{j}\right\rangle \right)  \right]
\\
\left\vert \omega^{\pm}\right\rangle _{ij}=\frac{1}{2}\left[  \left\vert
V\right\rangle _{i}\left(  \left\vert a_{j}\right\rangle +\left\vert
b_{j}\right\rangle \right)  \pm\left\vert H\right\rangle _{i}\left(
\left\vert a_{j}\right\rangle -\left\vert b_{j}\right\rangle \right)  \right]
\end{array}
.\right. \label{type2}%
\end{equation}
The Bell states in the two complementary bases satisfy the following
relations
\begin{equation}
\left\{
\begin{array}
[c]{c}%
\left\vert \Phi^{\pm}\right\rangle _{ij}=\frac{1}{\sqrt{2}}\left(  \left\vert
\chi^{\mp}\right\rangle _{ij}\pm\left\vert \omega^{\pm}\right\rangle
_{ij}\right) \\
\left\vert \Psi^{\pm}\right\rangle _{ij}=\frac{1}{\sqrt{2}}\left(  \left\vert
\chi^{\pm}\right\rangle _{ij}\pm\left\vert \omega^{\mp}\right\rangle
_{ij}\right)
\end{array}
\right.  ,\label{relation}%
\end{equation}
where $\left\vert \cdot\right\rangle _{A1}$ and $\left\vert \cdot\right\rangle
_{B2}$ represent the quantum states to be measured by Alice and Bob,
respectively. Because the Bell states are polarization and path entangled, the
complete Bell-state measurement can be done with only linear optics
\cite{martini}. This feature is essential to our deterministic quantum
cryptography protocol. After the measurements have taken place, Alice and Bob
can announce in public which type of Bell bases they have used, and then
divide the measurements into two sub-groups: For the first group they use the
same Bell bases, and for the second group they use the different Bell bases.
Of course, owing to the limited detection efficiency they only consider those
events in which both Alice and Bob successfully register a single photon.

Let us first consider what will happen when Alice and Bob use the same Bell
bases, i.e., they use either type-I \textit{or} type-II Bell base. In this
case the state $\left\vert \Psi\right\rangle _{A1B2}$ of Alice and Bob,
according to entanglement swapping, can be rewritten in the two Bell bases as
\begin{align}
\left\vert \Psi\right\rangle _{A1B2}  & =\frac{1}{2}\left(  \left\vert
\Phi^{+}\right\rangle _{A1}\left\vert \Phi^{+}\right\rangle _{B2}-\left\vert
\Phi^{-}\right\rangle _{A1}\left\vert \Phi^{-}\right\rangle _{B2}\right.
\nonumber\\
& \left.  -\left\vert \Psi^{+}\right\rangle _{A1}\left\vert \Psi
^{+}\right\rangle _{B2}+\left\vert \Psi^{-}\right\rangle _{A1}\left\vert
\Psi^{-}\right\rangle _{B2}\right) \nonumber\\
& =\frac{1}{2}\left(  -\left\vert \chi^{+}\right\rangle _{A1}\left\vert
\chi^{+}\right\rangle _{B2}+\left\vert \chi^{-}\right\rangle _{A1}\left\vert
\chi^{-}\right\rangle _{B2}\right. \nonumber\\
& \left.  +\left\vert \omega^{+}\right\rangle _{A1}\left\vert \omega
^{+}\right\rangle _{B2}-\left\vert \omega^{-}\right\rangle _{A1}\left\vert
\omega^{-}\right\rangle _{B2}\right)  .\label{swap}%
\end{align}
Eq. (5) implies that, whenever Alice obtain a measurement result, Bob will
always obtain the same result as long as his measurement is performed in the
same Bell bases. Therefore, perfect correlations can be built for those cases
in which both of them perform a measurement on their own particle in the same
Bell bases.

If Alice and Bob use different Bell bases (i.e., type-I \textit{and} type-II
Bell bases), the total state of Alice and Bob can be rewritten as either
\begin{align}
\left\vert \Psi\right\rangle _{A1B2}  & =\frac{1}{2\sqrt{2}}\left(  \left\vert
\Phi^{+}\right\rangle _{A1}\left\vert \omega^{+}\right\rangle _{B2}+\left\vert
\Phi^{+}\right\rangle _{A1}\left\vert \chi^{-}\right\rangle _{B2}\right.
\nonumber\\
& +\left\vert \Phi^{-}\right\rangle _{A1}\left\vert \omega^{-}\right\rangle
_{B2}-\left\vert \Phi^{-}\right\rangle _{A1}\left\vert \chi^{+}\right\rangle
_{B2}\nonumber\\
& -\left\vert \Psi^{+}\right\rangle _{A1}\left\vert \chi^{+}\right\rangle
_{B2}-\left\vert \Psi^{+}\right\rangle _{A1}\left\vert \omega^{-}\right\rangle
_{B2}\nonumber\\
& \left.  +\left\vert \Psi^{-}\right\rangle _{A1}\left\vert \chi
^{-}\right\rangle _{B2}-\left\vert \Psi^{-}\right\rangle _{A1}\left\vert
\omega^{+}\right\rangle _{B2}\right)  ,\label{a1}%
\end{align}
or
\begin{align}
\left\vert \Psi\right\rangle _{A1B2}  & =\frac{1}{2\sqrt{2}}\left(  \left\vert
\Phi^{+}\right\rangle _{B2}\left\vert \omega^{+}\right\rangle _{A1}+\left\vert
\Phi^{+}\right\rangle _{B2}\left\vert \chi^{-}\right\rangle _{A1}\right.
\nonumber\\
& +\left\vert \Phi^{-}\right\rangle _{B2}\left\vert \omega^{-}\right\rangle
_{A1}-\left\vert \Phi^{-}\right\rangle _{B2}\left\vert \chi^{+}\right\rangle
_{A1}\nonumber\\
& -\left\vert \Psi^{+}\right\rangle _{B2}\left\vert \chi^{+}\right\rangle
_{A1}-\left\vert \Psi^{+}\right\rangle _{B2}\left\vert \omega^{-}\right\rangle
_{A1}\nonumber\\
& \left.  +\left\vert \Psi^{-}\right\rangle _{B2}\left\vert \chi
^{-}\right\rangle _{A1}-\left\vert \Psi^{-}\right\rangle _{B2}\left\vert
\omega^{+}\right\rangle _{A1}\right)  .\label{a2}%
\end{align}
The above two equations imply the following: Any specific outcome obtained by
Alice or Bob in one of the two bases is maximally random and as such, given
two specific outcomes of measurements in one type of the Bell bases obtained
by Alice, she can predict with certainty the two outcomes of the corresponding
measurements performed by Bob in another Bell bases and \textit{vice versa}.
For example, whenever Alice obtains one of the two Bell states $\left\vert
\Phi^{+}\right\rangle $ and $\left\vert \Psi^{-}\right\rangle $ (or
$\left\vert \Phi^{-}\right\rangle $ and $\left\vert \Psi^{+}\right\rangle )$,
Bob's photon will be in one of the two states $\left\vert \omega
^{+}\right\rangle $ and $\left\vert \chi^{-}\right\rangle $ (or $\left\vert
\omega^{-}\right\rangle $ and $\left\vert \chi^{+}\right\rangle $). Thus,
Alice and Bob can always build two kinds of correlations in different bases.

Since a deterministic correlation can be built for every joint measurement,
all outcomes of the measurements in the same or different bases can be further
used as key bits. This is the essential property of our deterministic quantum
cryptography protocol and is different from the existing schemes with
two-dimensional entangled photon pairs.

Having established that deterministic correlations can be built between two
communicating parties, let us show how to generate the key bits in the present
protocol. First, if Alice and Bob have used the different bases, they can then
simply define the two kinds of correlations as $0$ and $1$. For example, in
the case of that Alice performs a measurement in the $\left\{  \left\vert
\Phi^{\pm}\right\rangle _{ij},\left\vert \Psi^{\pm}\right\rangle
_{ij}\right\}  $ bases and Bob performs a measurement in the $\left\{
\left\vert \chi^{\pm}\right\rangle _{ij},\left\vert \omega^{\pm}\right\rangle
_{ij}\right\}  $ bases, Alice can define the two Bell states $\left\vert
\Phi^{+}\right\rangle $ and $\left\vert \Psi^{-}\right\rangle $ as 0 and
$\left\vert \Phi^{-}\right\rangle $ and $\left\vert \Psi^{+}\right\rangle $ as
1, while Bob can define the two states $\left\vert \omega^{+}\right\rangle $
and $\left\vert \chi^{-}\right\rangle $ as 0 and $\left\vert \omega
^{-}\right\rangle $ and $\left\vert \chi^{+}\right\rangle $ as 1. Thus in this
case, one key bit can be built. Second, if they use the same Bell bases, they
actually receive two bits of deterministic information from the measurements
since there are $4$ possible outcomes of measurements in any one of the two
Bell bases. They can agree with each other in advance that the Bell states
($\left\vert \Phi^{+}\right\rangle $, $\left\vert \Phi^{-}\right\rangle $,
$\left\vert \Psi^{+}\right\rangle $, $\left\vert \Psi^{-}\right\rangle $) and
($\left\vert \chi^{+}\right\rangle $, $\left\vert \chi^{-}\right\rangle $,
$\left\vert \omega^{+}\right\rangle $, $\left\vert \omega^{-}\right\rangle $)
are encoded as two bits $00$, $01$, $10$ and $11$. Thus in the latter case,
two key bits can be built. Since the probabilities that Alice and Bob perform
the mesaurement in different Bell bases and in the same Bell bases are the
same, i.e. $50\%$, every measurement in which both of Alice and Bob register a
photon can establish $1.5$ key bits on average. Therefore, by exploiting the
high-capacity encoding method, the above scheme can provide a more efficient
quantum cryptography than the existing deterministic ones.

Let us now discuss the security of the scheme. For simplicity, suppose that an
eavesdropper, Eve, has full access to the quantum channel between Alice and
Bob, Eve can thus make use of intercept-resend attack. He can first randomly
perform a Bell-state measurement on the photon 2 in the two complementary
bases and then send its replacement to Bob. However, when Alice performs a
Bell-state measurement on her photon by randomly choosing one of the two
complementary bases, she equivalently prepares a mixture of some nonorthogonal
states with equal probability. Owing to the limit of the non-cloning theorem
\cite{woot}, it is not possible for Eve to know which basis Alice has chosen.
Indeed, Eve can at most get $50\%$ information, while Alice and Bob have about
$25\%$ error rate in the key bits obtained \cite{gis,ben1}. By announcing and
comparing in public some outcomes of the measurements in the same bases, Alice
and Bob can thus easily detect the presence of Eve.

However, Eve can also perform another type of intercept-resend attack. Eve can
first randomly perform a Bell-state measurement on both photons in one of the
two complementary bases and then send its replacements to Alice and Bob,
respectively. In this case, since any measurement performed by Eve will
destroy the original entanglement between the two photons, the measurement
results obtained by Alice and Bob during the verifictaion of entanglement
swapping will be inconsistent with the predictions of quantum mechanics.
Specifically, by comparing the outcome of the measurement in the same
Bell-state bases, in a single run Alice and Bob will detect the eavesdropper
with the probability of $1/4$\ whenever Eve performs the Bell measurements in
the same Bell bases and with the probability of $1/2$\ whenever Eve performs
the Bell measurements in the different bases. By announcing and comparing in
public some outcomes of the measurements in the same bases, again Alice and
Bob can easily detect the presence of Eve.

As pointed out in Ref. \cite{chen}, the state $\left\vert \Psi\right\rangle
_{A1B2}$ used in our protocol is effectively a maximally entangled state in a
4$\times4$-dimensional Hilbert space. Essentially, the scheme presented here
is thus a high-dimensional quantum cryptography protocol. In contrast to the
former high-dimensional quantum cryptography protocols where the protocols
provide better security against eavesdropping than the low-dimensional ones
while lead to a low generation rate of key bits \cite{bru,Bch1,Bch2,Bch3,Bour}%
, our protocol not only provides better security against eavesdropping but
also has a higher generation rate of key bits than the low-dimensional ones.

Compared with Ekert's quantum cryptography protocol using EPR-entangled qubit
states \cite{ekert}, a striking feature of the present scheme is its higher
efficiency. In Ekert's scheme, in order to be against eavesdropping, Alice and
Bob randomly perform statistical measurements and register the outcomes in one
of three nonorthogonal bases. When they have used the same bases incidentally,
the results are correlated or anticorrelated and can be converted into key
bits. For the others the results is not correlated and therefore used only for
checking whether there is eavesdropper with help of Bell's inequalities. Thus,
the overall probability of creating a key bit is $2/9$ per pair. However, in
our scheme each measurement in which both of Alice and Bob register a photon
simultaneously can generate $1.5$ key bits on average. Thus, the key bit
generation efficiency per pair is $27/4$ times higher in our scheme than in Akert's.

In summary, we have presented a protocol for deterministic and efficient
quantum cryptography with two photons that are entangled in both polarization
and path degrees of freedom. The protocol can establish $1.5$ key bits per
entangled pair, which is almost one order of magnitude ($\sim7$ times) higher
than the Akert's scheme and is completely secure against eavesdropping. The
polarization- and path-entangled photon pairs \cite{pan2} and the complete
Bell-state measurement \cite{martini} are two basic ingredients of the
proposed protocol. Thus, our scheme is feasible and within the reach of
current technology. The experimental realization of the protocol is under way
in our lab.

\begin{acknowledgments}
This work was supported by the National Natural Science Foundation of China,
Chinese Academy of Sciences and the National Fundamental Research Program (
under Grant No. 2001CB309303 ).
\end{acknowledgments}

\end{document}